\documentclass[aps,twocolumn,reprint,superscriptaddress]{revtex4-1}
\usepackage{amsmath}
\usepackage{epsfig}
\usepackage{graphicx}
\usepackage{graphicx, color, epstopdf}
\usepackage{bm}
\usepackage{amssymb}
\usepackage{xcolor}
\usepackage{subfigure}
\usepackage[colorlinks, linkcolor=red, anchorcolor=green,citecolor=blue ]{hyperref}

\begin{document}
\bibliographystyle{apsrev4-1}

\title{  Dynamic evolution  of S$_0$-S$_3$ at the oxygen evolving complex with spin markers under photoelectric polarization   }
\author{Pei-Ying Huo}
\affiliation{School of Physics, Southeast University, Nanjing 211189, China}

\author{Wei-Zhou Jiang}
\email{Corresponding author:wzjiang@seu.edu.cn}
\affiliation{School of Physics, Southeast University, Nanjing 211189, China}

\author{Rong-Yao Yang}
\affiliation{School of Physics, Southeast University, Nanjing 211189, China}

\author{Xiu-Rong Zhang}
\affiliation{School of Science, Jiangsu University of Science and Technology, Zhenjiang 212100, China}

\begin{abstract}
 Water oxidation at the oxygen evolving complex (OEC) of the photosystem II is catalyzed by the core cluster CaMn$_4$O$_5$ which  was projected to experience five intermediate states S$\rm_i$ in the Kok's cycle since 1970's. However, the detailed dynamics of  state evolutions still remains unclear, albeit with the general fact that the process is initiated by  the transfer of photoelectrons with the steady electron donors of the water molecules. Based on density functional simulations, we find that the spin flips of Manganese atoms between the consecutive states  in the electric polarization field can be used as a marker to uncover the intricate dynamic evolutions and the underlying dynamics.   The dynamic electron and proton transfers and water insertion and dissociation are traced to reveal the evolution pathways  of S$_0$-S$_3$ with commensurate spin flips  towards the exact spin configuration of the next state. In particular, the various water insertions and dissociations at coordination sites of the S$_2$ open and closed cubane isomers are predicted with constraints on the necessary spin flips. Our study paves a way to uncover the animated OEC evolutions with the spin flips that await for  more experimental verifications and lays a solid ground for revealing the mechanism of dioxygen generation via the pending S$_4$ state.
\end{abstract}

\maketitle

\section{INTRODUCTION}
In nature, the solar energy is converted into the  biochemical energy in the  photosynthetic process that enables the self-sustained
biological circulation on the Earth~\cite{shen2015structure}.
In the cascade photosynthetic process, one of the most fundamental reactions is the water oxidation catalyzed at the oxygen evolving complex (OEC).
Not only does the water oxidation release  the molecular oxygen to refresh the atmosphere vitally in the industrial age, but provides the logistic supply of 4 electrons and 4 protons in a closed circle  to convert carbon dioxide into the organic molecules. The catalytic pathway of the OEC is known as the Kok's cycle~\cite{kok1970cooperation} that   circulates
through five redox states ($\rm S_i, i=0,\cdots4,$), driven by the carousel of
photoelectron in the Photosystem II (PSII) reaction centre, P$_{680}$. In the past, great endeavor both in experiments and simulations has been made to identify the structures of OEC states $\rm S_{i=0,\cdots,3}$ prior to the dioxygen generation at ${\rm S}_4$ whose structure was reported most recently~\cite{greife2023electron}. Since the mechanism of molecular oxygen formation still remains rather elusive, the standing effort to reveal the OEC structures and dynamic evolutions is important not only for understanding the nature of water oxidation but also for  designing  the catalysts to develop clean
energy~\cite{zhang2015synthetic, zhang2018nature, wilson2018structural}.

Since the OEC core structure in the dark stable state (S$_1$) was first revealed by high resolution X-ray diffraction in 2011~\cite{umena2011crystal},  experimental and computational efforts  have further been made to achieve outstanding progresses such as refining respective $\rm S_i$ structures~\cite{suga2015native, askerka2015analysis, pantazis2012two, young2016structure, suga2017light, kern2018structures, pal2013s0, li2015simulation, shen2015structure}, determining oxidation states~\cite{krewald2015metal, cox2014electronic, petrie2015resolving, terrett2016computational,
baituti2018computational, schuth2018kalpha}, and spin orientations~\cite{bovi2014magnetic, krewald2015metal, krewald2016spin, chatterjee2016structural} of Mn atoms. However, the evolution
pathway of the OEC states prior to the dioxygen generation, especially S$_2$$\rightarrow$S$_3$ that involves water insertion, electron and proton transfers, still remains quite unclear in details. The S$_2$ state features two EPR signals with  g=2 and  4.1 which were  identified by computational studies to be the low-spin open and high-spin closed cubane isomers (S$_2$-a and S$_2$-b), respectively~\cite{pantazis2012two}. The evolution of S$_2$$\rightarrow$S$_3$ may thus suggest several possible paths, with variations arising from the  substrate water molecules (W2 or W3), the function of two isomers and the exact sequence of the three events~\cite{allgower2022molecular, bovi2013s2, shoji2015qm, ugur2016redox, askerka2015nh3, retegan2016five, capone2015reorganization, capone2016mechanism}, and the transition from S$_2$-a to S$_2$-b in the evolution path was once regarded as a key mechanism to obtain the high spin (S=3) of the S$_3$ state~\cite{krewald2016spin}. Recently,  the insertion of water in the S$_2$ open cubane isomer in the evolution process has gained mounting support~\cite{ibrahim2020untangling, siegbahn2018s, allgower2022molecular, hussein2021structural}. It was reported that the protonation of O$_4$ in S$_2$-a can induce a low-spin to high-spin transition from 1/2 to 5/2 (or even 7/2) without structural isomerization~\cite{corry2019proton, corry2020molecular}.  Actually, the detail how the spin of various states evolves and the dynamics that dictates the spin flip remain unclear to large extent due to the complexity subject to the dynamic complex system.  These issues are not limited to S$_2$$\rightarrow$S$_3$ but pervasive in the evolution of the whole Kok's cycle.

The state evolution is triggered by the polarization effect of the positive-charged hole after the transfer of the photoproduced electron at P$_{680}$.  The hole polarization field  drives the electron transfer from the core Mn$_4$Ca that breaks the time reversal symmetry associated with the spin flips of the state.
The spin flips can thus reflect an irreversible time sequence of the evolution.  More importantly, the spin flip can serve as a significant indicator to reveal the detail of the dynamic evolution, because it is associated tightly  with the variation of the polarization field and the resultant charge transfers.
In fact, the electric polarization provides a gradient field that modifies the Coulomb potential, while the change in the Coulomb potential can  shift  the spin or orbital motion, in that the spin-orbit coupling potential is determined by the gradient of the Coulomb potential according to the Dirac equation of charged particles. In addition, the bonding electrons in molecular orbitals,  restricted by the Pauli exclusion principle, are characterized by their spin states in the Coulomb potential. Thus, the electric polarization should naturally result in the shift of the spin dynamics of atoms in the OEC. In this work, we study for the first time the polarization effect on the OEC cycle. In particular, with the density functional simulations, we explore the spin flips of Manganese atoms and reveal the dynamic mechanisms of state evolutions of S$_0$-S$_3$, driven by the polarization of the photoelectric hole at  P$_{680}$.

\section{Computational Methods}

\begin{figure}[thb]
\centering
\includegraphics[width=0.85\columnwidth]{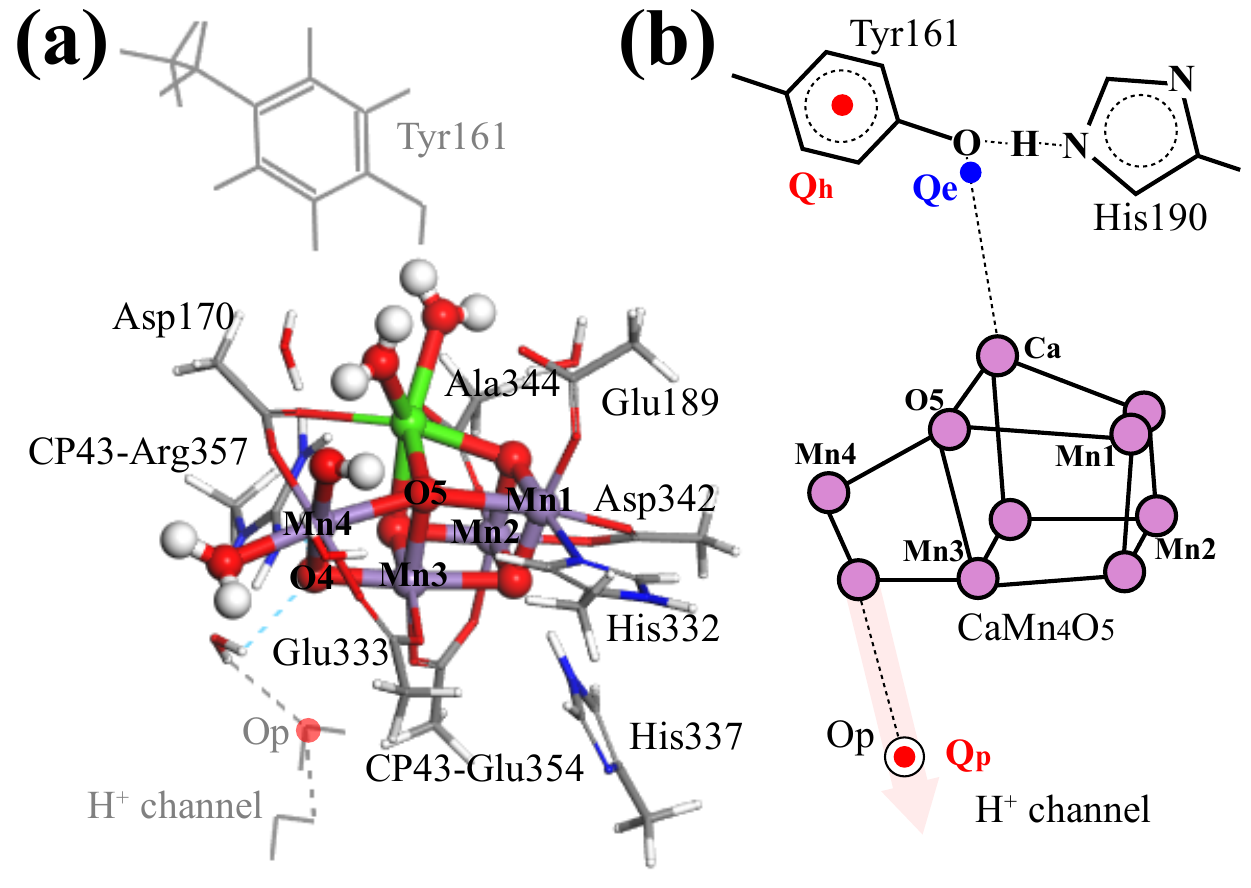}\\
\caption{(a) Geometric model of the OEC CaMn$_4$O$_5$, along with surrounding  ligated amino acid residues. Purple, green, red, grey, blue and white spheres represent Mn, Ca, O, C, N, H atoms, respectively.  The Tyr161 outside of the OEC and  one proton transfer channel, denoted by $\rm O_p$, are also sketched.  (b) Schematic diagram for the positions of effective point charges set in the simulation. The red and blue solid circles represent positive and negative point charges, respectively.
}\label{fig1}
\end{figure}

We perform the spin-polarized all-electron density functional simulations with the DMol$^3$ package in the Materials Studio of Accelrys Inc. The OEC system in the simulation is intercepted from the XRD crystal data obtained at 1.9~\AA~resolution by Yasufumi Umena\cite{umena2011crystal} from protein data bank (PDB ID: 3WU2). As shown in the Fig.~\ref{fig1}a, the selected configuration is composed of the core structure (CaMn$_4$O$_5$) and its nine surrounding amino acids (His332, Glu189, Asp342, Ala344, Asp170, Gul333, CP43-Glu354, His337, CP43-Arg357), as well as eight water molecules. Eventually, the system contains 112 atoms. In the geometric optimization process, we adopt the  Perdew-Burke-Ernzerhof (pbe) exchange correlation functional within the general gradient approximation (GGA) and the double numerical basis set plus polarization (DNP). Optimization accuracy of the energy is set to be $2.0\times10^{-5}$ Hartree. All eight possible Mn spin configurations of CaMn$_4$O$_5$ structure are considered during the optimization procedure and  the lowest energies are regarded as the S$\rm_i$ structures. Mn oxidation states are calculated by both the Mulliken spin populations and the bond valence sum (BVS). In the BVS calculation, relevant parameters are taken from the Ref.~\cite{liu1993bond}.

The hole polarization effect following each light-driven electron separation at P$_{680}$ is simulated by putting an effective point charge (denoted as $Q_h$) at the tyrosine residue (Tyr161). Electron separation at P$_{680}$ leads to oxidation of Tyr161, which subsequently oxidizes the core CaMn$_4$O$_5$. Note that some computational calculations and experiments suggested that the Tyr161 oxidation is followed by a reciprocal deprotonation toward His190 and leaves a neutral radical~\cite{nakamura2020pivotal, nakamura2014fourier, styring2012two}. The proton migrates back to Tyr161 with a cooperative electron transfer from the CaMn$_4$O$_5$. This is actually a rather complicated process, while we make a simple comparison to the case with  a positive point charge put at His190. It is found that the resulting spin flips and system energies are consistent for the electron transfer.  Additionally, simulations confirm the negligible difference in the  impact of the specific location of the hole  at the central or peripheral C or O atoms of Tyr161. Based on these verifications, we eventually place  Q$_h$ at the central site of Tyr161. Considering the distance and screening effects, the effective charges of $Q_h\approx0.1$e and  $Q_h$=1.0e at the Tyr161 correspond to the photoelectric hole generation at P$_{680}$ and  Tyr161 oxidization, respectively. Intrigued by the hole polarization effect, the electron and proton are deprived sequentially from the core CaMn$_4$O$_5$ cluster. During deprivations of electrons and protons along e$^-$/H$^+$ transfer channels, the polarization effect is also  produced by the moving  electrons ($Q_e$) and protons ($Q_p$). The polarization effects by both the hole and moving electron and proton are simulated by putting  the equivalent point charges at some typical places under the constraint of the charge conservation of the whole system to track the dynamic process of state evolution. The point charge for simulating the moving electron ($Q_e$) is  placed close to Tyr161 and typically at a distance of 4~\AA~ away from Ca atom of CaMn$_4$O$_5$ on the line connecting to the O atom of Tyr161, see Fig.~\ref{fig1}b. The specific choice of the charge site is partly  associated with the fact that the proton migrations with a cooperative electron transfer  take place near the O atom after  the Tyr161 oxidation. Meanwhile, it is verified that  varying Q$_e$'s location on the various lines pointing towards the O, C atoms and  centre of Tyr161  at a distance of 4~\AA ~away from the Ca atom of CaMn$_4$O$_5$ does not change the main results concerning the spin flips of Mn atoms.  The polarization effect of the moving proton is simulated by putting the equivalent point charge of 1.0e  at the site O$_p$ near the O atom along the H$^+$ transfer channel, as shown in Fig.~\ref{fig1}.

\section{Results and Discussion}
\subsection{Structures of S$_0$-S$_3$ states}
\begin{figure}[thb]
\centering
\includegraphics[width=0.65\columnwidth]{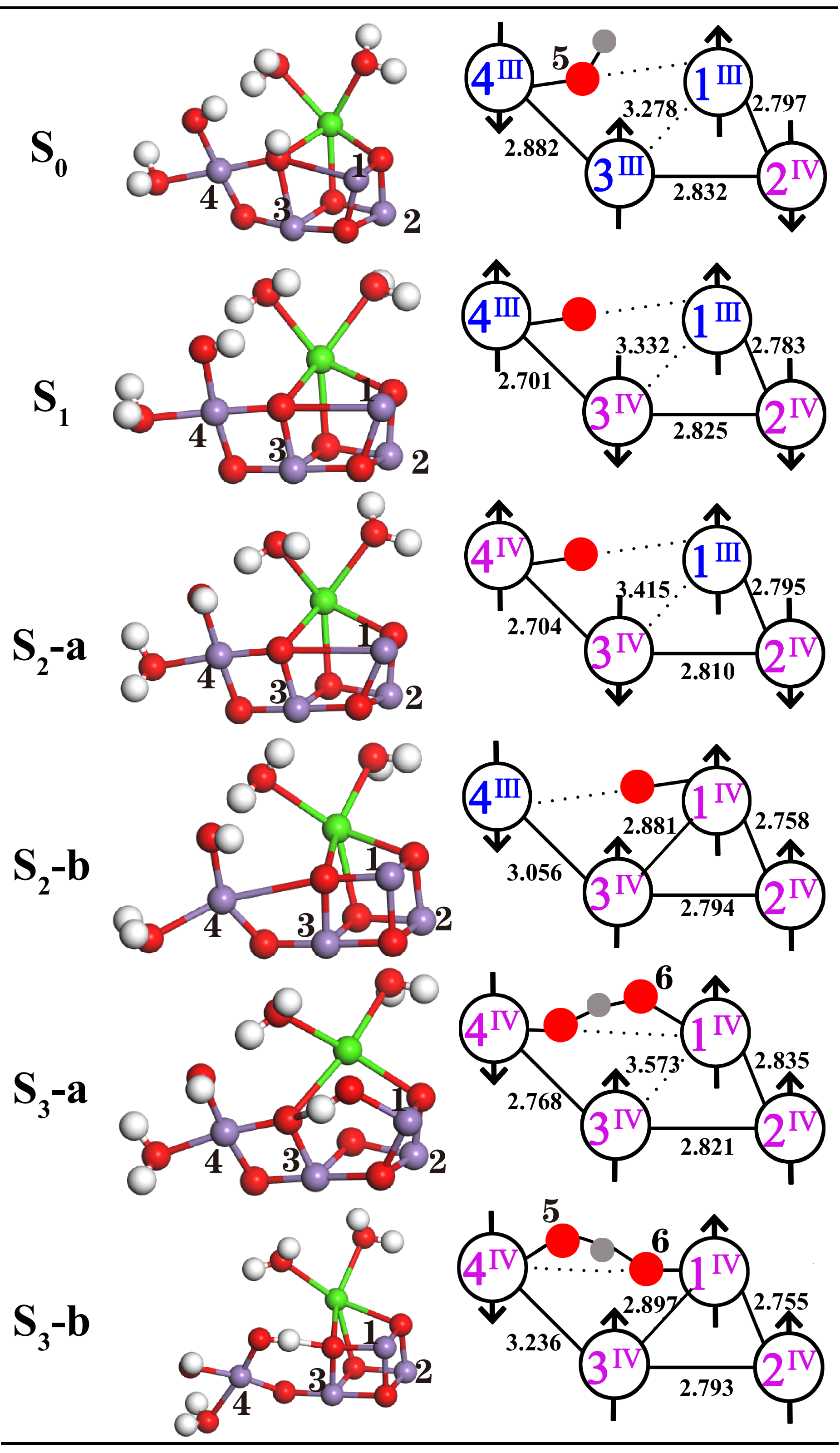}\\
\caption{Optimized structures of the S$_0$-S$_3$ states. Shown on the right column are Mn-Mn distances (\AA), spin orientations and oxidation states of Mn atoms.
}\label{fig2}
\end{figure}

In recent years, there has been a noteworthy progress in revealing the structural details of the states S$\rm_i, i=0,\cdots,3$. The Mn$_1$$\sim$Mn$_4$ oxidation states for S$_0$-S$_3$ states are suggested  by experimental and computational studies to be in turn (III, IV, III, III)$\rightarrow$(III, IV, IV, III)$\rightarrow$(III, IV, IV, IV)/(IV, IV, IV, III)$\rightarrow$(IV, IV, IV, IV) for S$_0\rightarrow$S$_1\rightarrow$S$_2$-a/S$_2$-b$\rightarrow$S$_3$~\cite{suga2015native, krewald2015metal, yachandra1993plants}.
Various measurements by the extended x-ray absorption fine structure (EXAFS) have revealed a consistent picture of the Mn-Mn distances of each S state~\cite{krewald2015metal}. The spin configurations of Mn atoms in  S$_0$-S$_3$ states are also identified:  $\uparrow\downarrow\uparrow\downarrow$ in the S$_0$ state in which one proton is assigned to the O$_5$ ~\cite{krewald2016spin},   $\uparrow\downarrow\downarrow\uparrow$ in the dark-stable  state S$_1$ with the ground state spin S$_{GS}$=0~\cite{koulougliotis1992oxygen, yamauchi1997parallel},   $\uparrow\downarrow\downarrow\uparrow$ and $\uparrow\uparrow\uparrow\downarrow$ for two S$_2$ isomers S$_2$-a and S$_2$-b, respectively~\cite{bovi2014magnetic, krewald2016spin}, and  $\uparrow\uparrow\uparrow\downarrow$ or $\uparrow\uparrow\downarrow\uparrow$  for S$_3$-a and $\uparrow\uparrow\uparrow\downarrow$ for S$_3$-b~\cite{krewald2015metal, krewald2016spin}. Note that there are some uncertainties in S$_3$ structure arising from three measurements by the X-ray free electron laser (XFEL) during 2016-2018~\cite{young2016structure, suga2017light, kern2018structures}. It was reported in 2016  that no additional water or hydroxo bond to the OEC was observed in S$_3$~\cite{young2016structure}, whereas other two measurements confirmed the presence of an additional oxygen ligand but with different O$_5$-O$_6$ (Ox) distances~\cite{suga2017light, kern2018structures}. The simulations show there exist two S$_3$ isomers, S$_3$-a and S$_3$-b, obtained from S$_2$-a and S$_2$-b by adding one hydroxyl, respectively~\cite{retegan2016five, pantazis2019s3}.  Our optimized structures for S$_0$-S$_3$ including the Mn-Mn bond distances, oxidation states and spin configurations of Mn atoms are displayed in  Fig.~\ref{fig2}, which are consistent with those in the literature mentioned  above.

\subsection{Dynamic evolutions of S states}

Prior to the analysis of  the specific  dynamic evolution, it is necessary to understand the force with the underlying potential that drives the evolution of S states. The OEC cycle starts with the production of the photoelectric hole after the photoelectron separation, which exerts an electric polarization effect on Tyr161 and  OEC. The presence of the polarization effect naturally adds  the modification to the Coulomb field gradient appearing in the second-order Dirac  equation of charged particles. The gradient term is given as
\begin{equation}\label{eqso}
\xi(r)=-\frac{1}{2m_e^2c^2}\vec{\sigma}\cdot({\bf \hat{p}}\times\nabla V),
\end{equation}
where $V$ is the Coulomb potential and $\vec{\sigma}$  the spin Pauli matrices. In the central force,  Eq.(\ref{eqso}) just represents the spin-orbit coupling term: $\xi(r)=\vec{\sigma}\cdot{\bf \hat{L}}/(2m_e^2 r c^2)dV/dr,$ with  the angular momentum operator ${\bf \hat{L}}$ obtained from the curl calculation.  In the specific density functional, the contribution of $\xi(r)$ should be included appropriately in the potential terms of the Kohn-Sham equations. The modification of the Coulomb field gradient can thus shift the  spin orientations of unpaired electrons and affect the electron orbitals. The spin states of atoms, restricted by the Pauli exclusion principle, are  informative  for the bonding  properties and orbital motions of the molecule due to the spin-orbit interaction.   Also, the electron spins on the hybrid orbitals can also reflect the structural  properties including electron  spatial configuration of molecules. While different spin states of atoms correspond definitely to different quantum states and valence bond properties, the spin flips  under the polarization field of the positive-charged hole and moving electrons and protons can be used as a marker to track the dynamic evolution of the S state, providing a theoretical criterion to determine the controversial evolution pathways.

\begin{figure}[thb]
\centering
\includegraphics[width=0.75\columnwidth]{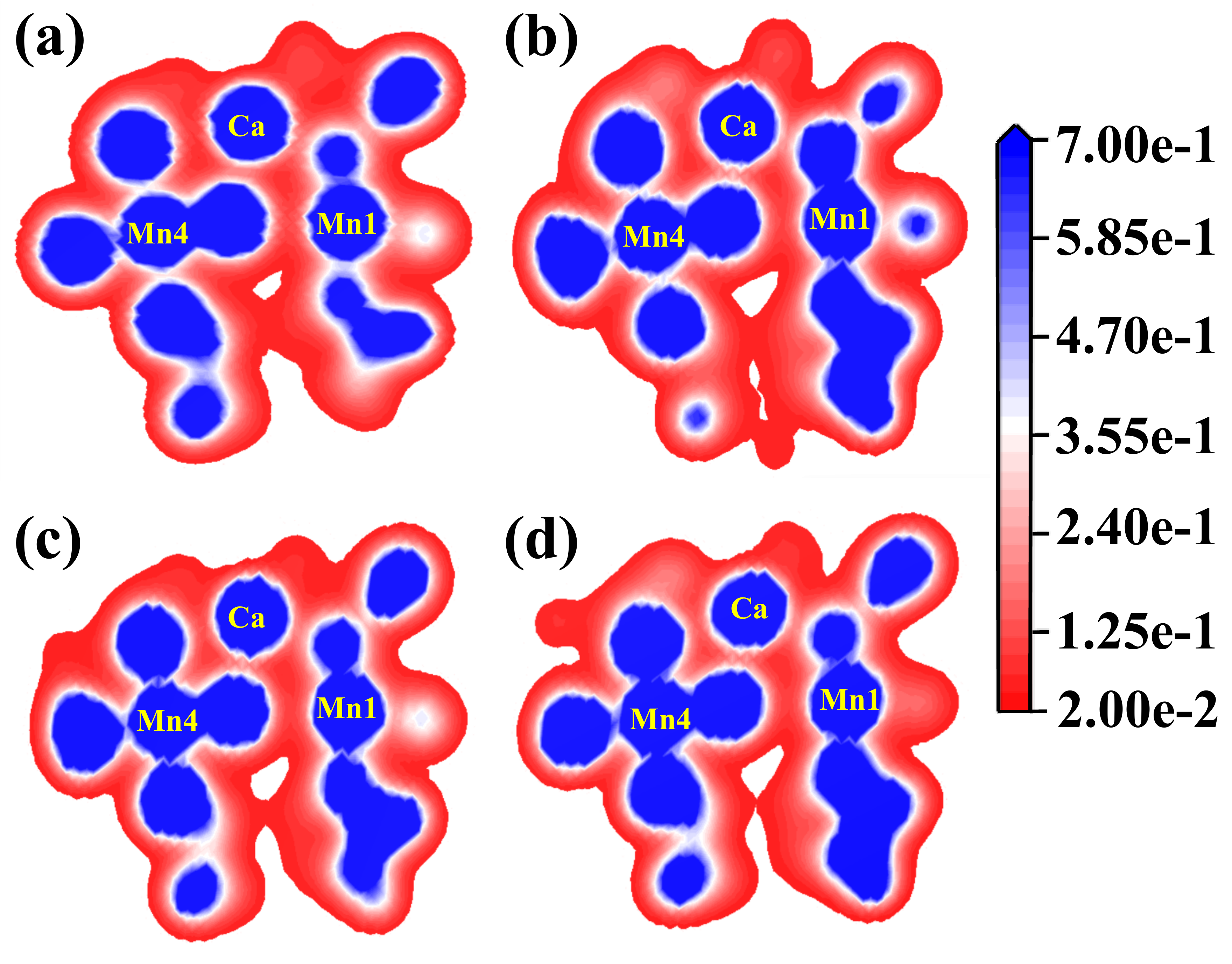}\\
\caption{Sectional profiles of the electron density distributions at the Ca-Mn$_1$-Mn$_4$ plane}. Slices (a), (b), (c) and (d)  stand for the cases for  the  pure S$_1$ state,  with the hole at the Tyr161, with the transferring effective charge Q$_e$ of -0.5e and -0.9e  4~\AA~away from the Ca atom of CaMn$_4$O$_5$, respectively.
\label{fig3}
\end{figure}

As the potential in the density functional equations is modified by the polarization field, the obtained electron wave functions and density distributions undergo the change accordingly. Here we present an example of the electron  distribution variations of the S$_1$ state in the polarization field produced by the electron hole at the Tyr161 and the  electron transfer. Shown in Fig.~\ref{fig3} are the sectional electron density distributions of the S$_1$ state in various cases.  We can observe some significant variations in the electron density distributions and  geometric structures arising from the various  polarization effects. When the hole polarization is imposed on the $\rm S_1$ state, the striking changes in the density profiles  can be observed, as compared with the right and lower regions of Fig.~\ref{fig3}a,b. On the lower part of Fig.~\ref{fig3}a,b, the relative change arises mainly from the tilting and prostrating of the amino acid residues induced by the polarization field.  As the charge of the moving electron increases,  the significant changes also occur at edge parts including the amino acid residues, see the relative changes in Fig.~\ref{fig3}a,b,c.  As these variations are induced by the underlying changes in the Coulomb potential that are associated with the spin-space (orbit) coupling,  as seen in Eq.(\ref{eqso}),   using the spin as a marker can point to the intermediate state structures that are characterized by the spin of atoms, electron density distributions and geometric features consistently  in the dynamic evolution intrigued by the polarization field.

\textbf{S$_0\rightarrow$S$_1$.}  The hole polarization field  at P$_{680}$ leads to oxidation of Tyr161 and subsequently the CaMn$_4$O$_5$. At first, we employ a simple model without Tyr161 oxidation to simulate the hole polarization field which is numerically realized by placing an equivalent positive charge of $0.1$e at the Tyr161 considering the screening effect at a long distance. On the way of the electron transfer towards the hole at P$_{680}$, the effective charge of the hole reduces for the charge neutralization, and this is simulated by reducing the equivalent charge at the Tyr161 under the constraint of the charge conservation of the whole system.  After the simulation of this simple model, we consider the more realistic process of Tyr161 oxidation that  occurs in a time scale of the nanoseconds to microseconds and the CaMn$_4$O$_5$ oxidation in the time scale of the microseconds to milliseconds~\cite{styring2012two, brettel1984nanosecond, christen1998origin, rappaport1994kinetics, haumann2005photosynthetic}. After the oxidation of Tyr161 and the subsequent charge annihilation at the P$_{680}$, the hole polarization field at the Tyr161 dominates the oxidation of  the CaMn$_4$O$_5$. In this case, the positive charge of  $Q_h$=e is  placed at the Tyr161 to simulate the dynamic oxidation process and the evolution from the S$_0$ to S$_1$. We mimic the pathway of the electron transfer from the CaMn$_4$O$_5$ to the Tyr161 dynamically by guaranteeing correct spin flips of the Mn atoms from the $\rm S_0$ to $\rm S_1$.

The presence of the hole polarization with the equivalent Q$_h$ of 0.1e  point charge  at the Tyr161 causes the Mn$_4$ spin of S$_0$ state flips  surely due to the change  of the gradient of the Coulomb field. And the total energy increases under the hole polarization field, suggesting the higher  structural activity that is more suitable for the state evolution. Driven by the hole polarization, one electron and subsequently one proton start to be deprived from the CaMn$_4$O$_5$  for the evolution S$_0$$\rightarrow$S$_1$~\cite{klauss2012alternating}. In a simulation that  the moving electron approaches the hole  with the gradual Q$_h$ decrease  starting from 0.1e, the Mn spin configuration remains $\uparrow\downarrow\uparrow\downarrow$, whereas the Mn$_3$-O$_5$ distance gradually decreases to 1.975~\AA. Since the  Mn oxidation state depends on the proximity with the surrounding O atoms, the shortening of Mn$_3$-O$_5$ distance accounts for the transition of the oxidation state of Mn$_3$ to IV, which is supported by both Mulliken population and BVS calculations. Subsequently, the proton assigned to the O$_5$ is transferred, and the Mn spin configuration changes to $\uparrow\downarrow\downarrow\uparrow$, which is exactly in accordance with the spin of S$_1$ state.

\begin{figure}[thb]
\centering
\includegraphics[width=0.8\columnwidth]{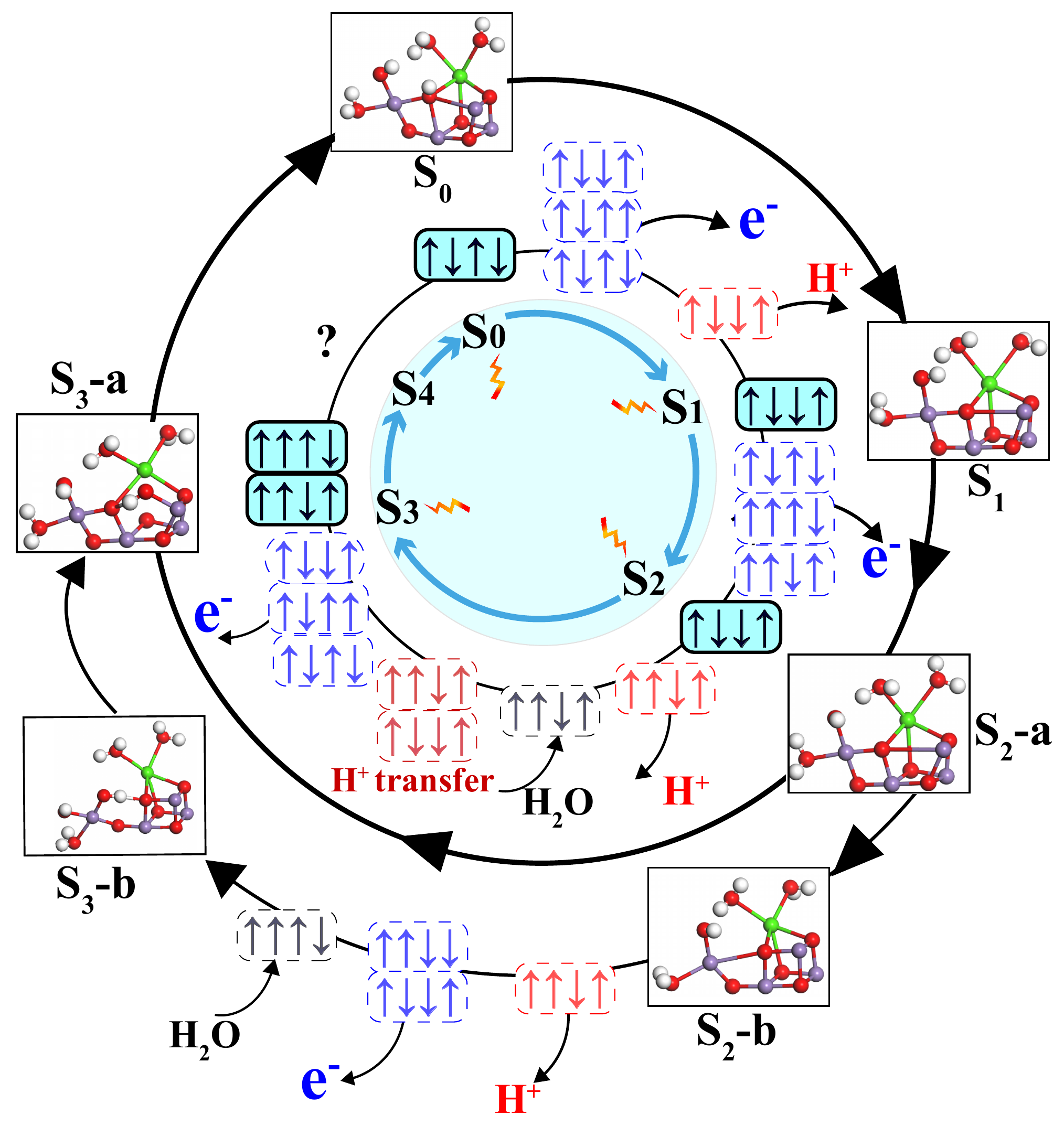}\\
\caption{Schematic diagram of the S$_i$ state cycle with  the spin markers. The blue and red dashed boxes show the transient Mn spin configurations during electron and proton transfer under the hole polarization.
}\label{fig4}
\end{figure}

Besides the simulation in a simple model, we investigate the   evolution  that is driven by the  oxidation of the positive-charged hole at the Tyr161.  The simulation with  the hole polarization field, induced by Tyr161 oxidation, also successfully reveals the spin flips of the Mn atoms in processes of the electron and proton deprivation and transfer. The Mn spin configuration changes from $\uparrow\downarrow\uparrow\downarrow$ to $\uparrow\downarrow\downarrow\uparrow$ under the hole polarization field. Owing to the close relationship between the spins and valence bond properties, the spin flips of the key atoms means the structural change of the molecule that puts the OEC state on a track of the dynamic evolution. It is found that the distances of some key bonds undergo significant changes in the hole polarization field. For instance, we notice that the  Mn$_3$-O$_5$ distance elongates from 2.263~\AA~to~ 2.344~\AA, compared to the slight changes~($\textless$0.02~\AA~) in other Mn-O distances,  which implies the weakening of the coupling strength between Mn$_3$ and O$_5$.  As the coupling is specified by the equivalent charge of the bonding atoms, the changes in the equivalent charge and bonding distances can play a role of the  precursor in  the electron transfer in the dynamic evolution.  To mimic the dynamic process of the electron transfer gradually towards the Tyr161 hole, we put an effective charge  Q$_e$  changing from -0.1e to -0.9e at the position  4~\AA~away from  the Ca atom of CaMn$_4$O$_5$ on the line connecting to the O atom of Tyr161 shown in Fig.~\ref{fig1}b. Consequently,  the polarization effect, produced by  the hole (Q$_h$=e) and moving electron,  diminishes and eventually vanishes when  the moving electron fully recombines with the hole. In this process, along with the Mn$_3$-O$_5$ distance decreasing, the Mn spin configuration goes through $\uparrow\downarrow\downarrow\uparrow$ and $\uparrow\downarrow\uparrow\uparrow$ and  back to $\uparrow\downarrow\uparrow\downarrow$ again, as shown in the blue dashed line box between the states S$_0$  and S$_1$ in  Fig.~\ref{fig4}. Then, the Mn spin configuration transits to that of the S$_1$ state after the proton removal.

In both cases, we see that the polarization effect of the hole directly causes the change of the bond lengths, the spin flips of Mn atoms, and the  OEC oxidation with the charge transfer.  Using the spin flip as a marker that points to the pathway of the moving electron and proton, we successfully illustrate the dynamic evolution of the state S$_0$ to S$_1$, which also provides  a paradigm for exploring the subsequent state evolutions by watching the spin flips of the Mn atoms.

\begin{figure*}[thb]
\centering
\includegraphics[width=1.6\columnwidth]{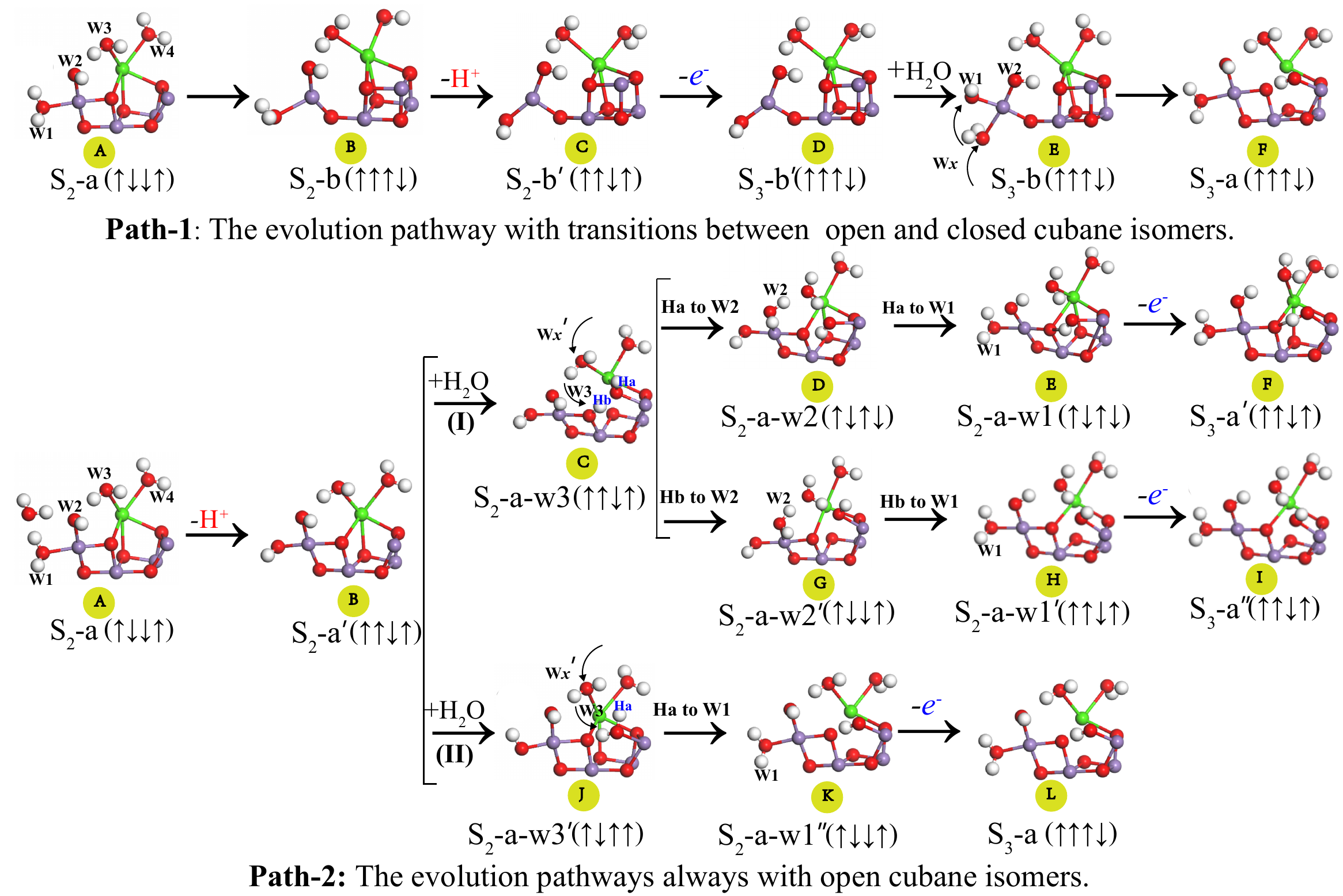}\\
\caption{ Evolution pathways for S$_2$$\rightarrow$S$_3$ with and without transitions between the open and closed cubane isomers, denoted as path-1 and path-2, respectively. The evolution includes the proton (-H$^+$) and electron (-$e^-$) transfers and the water insertion (+H$_2$O) as marked, and the states after the electron transfer are dubbed  $S_3$ states. W$x$ or W$x^{'}$ represents the ambient water on the insertion pathway. In path-2, branches (I) and (II) represent two different W3 insertion, and in path-2 (I), two H$^+$ from W3 in the diagram (C) are marked as Ha and Hb, respectively.}
\label{fig5}
\end{figure*}

$\textbf{S$_1\rightarrow$S$_2$.}$ A key object in understanding the S$_1\rightarrow$S$_2$ that involves only one electron transfer is to identify the terminal state in the presence of the two S$_2$ isomers. In the simple model with the effective charge Q$_h$ of 0.1e placed at the central site of Tyr161, the Mn spin configuration changes through $\uparrow\downarrow\uparrow\downarrow$  to  $\uparrow\downarrow\downarrow\uparrow$ during the electron transfer process as Q$_h$ gradually decreases from 0.1e. On the other hand, the  realistic hole polarization field produced by Tyr161 oxidation causes the Mn spin configuration to transit eventually to $\uparrow\downarrow\downarrow\uparrow$  through the intermediate states $\uparrow\downarrow\uparrow\downarrow$, $\uparrow\uparrow\uparrow\downarrow$  and $\uparrow\uparrow\downarrow\uparrow$, as the moving electron approaches the hole, as shown in the blue dashed boxes between the states S$_1$ and S$_2$ in Fig. \ref{fig4}. In both cases, the evolution of S$_1\rightarrow$S$_2$ reaches the ultimate Mn spin configuration of $\uparrow\downarrow\downarrow\uparrow$, which designates  the open-structure of S$_2$, the S$_2$-a isomer. During the evolution process, we observe that  the Mn$_1$-O$_5$ distance increases generally, which also supports that S$_2$-a is the terminal state of the S$_1$ evolution. Here, we stress that the spin marker identifies the terminal state S$_2$-a of the evolution,  while the  S$_2$-b  can be converted from S$_2$-a as two isomers are interconvertible~\cite{bovi2013s2, pantazis2012two}.

$\textbf{S$_2\rightarrow$S$_3$.}$ The S$_2$$\rightarrow$S$_3$ transition is a rather complex process that involves the insertion of water and the transfer of an electron and a proton. The exact sequence of these three events is not very clear yet, although various experimental measurements have confirmed that the proton removal takes place before the electron transfer~\cite{klauss2012alternating, ibrahim2020untangling, mausle2020activation, zaharieva2016room, takemoto2019proton, sakamoto2017monitoring}. Besides, it remains an hot issue to identify  the  pathway of the substrate water insertion and the specific roles of two S$_2$ isomers in the evolution process.

In the simulations, we consider all possible substrate water insertions, sequences of the electron and proton transfers,  and the participation of the two isomers.  Using the spin flip between the initial and terminal states as a marker, we can determine the candidate pathways of the evolution by ruling out others that are not compatible with the necessary spin flips. With the correct spin flips of Mn atoms,  dynamic evolution paths with and without transitions between open and closed cubane isomers are denoted as path-1 and path-2, respectively, as shown in Fig.~\ref{fig5}.

The path-1 starts from the S$_2$-a isomerization to S$_2$-b and then follows the sequence of the proton and electron transfers and water insertion. The proton transfer of W1 gives rise to structural rearrangement and opens an accessible coordination site for the cascade water insertions. As a result,  the hydroxyl (W2) is inserted between Mn$_4$ and O$_5$ in the intermediate state denoted  as S$_2$-b$^\prime$ in Fig.~\ref{fig5}.  At this moment, the Mn spin configuration changes to $\uparrow\uparrow\downarrow\uparrow$ as displayed in the simulations for both cases of the simple model and of the realistic hole  at the Tyr161.  The subsequent  electron transfer makes the  Mn spin configuration change to $\uparrow\uparrow\uparrow\downarrow$.  In a dynamic simulation through increasing the effective charge of the transferring electron Q$_e$, we can observe  that the Mn spin configuration undergoes  $\uparrow\uparrow\downarrow\downarrow$ and $\uparrow\downarrow\downarrow\uparrow$ till to  $\uparrow\uparrow\uparrow\downarrow$,  which reveals a dynamic process of the transient spin evolution.  Once the electron transferred from the OEC recombines with the hole, the polarization field vanishes until the  hole is regenerated in the next photoelectric process at P$_{680}$. Note that the principal spin flip from the low-spin S$_2$ to high-spin S$_3$ in path-1 is realized at the transition from the state S$_2$-a to S$_2$-b, similar to that in Refs.~\cite{bovi2013s2, retegan2016five, krewald2016spin}.  Then, the  water insertion can take place spontaneously. The experimental ammonia binding result and computational calculations claimed that a water molecule W$x$ around O$_4$, approaching close to Mn$_4$, causes the W1 and W2 to be displaced, see  the diagram (E) of Path-1 in Fig. \ref{fig5} ~\cite{askerka2015nh3, retegan2016five, capone2015reorganization, capone2016mechanism, wang2017crystallographic}. In our simulations, it shows that  the energy of the system decreases by 0.256~eV after the water insertion, which indicates that W$x$ as a substrate water is energetically favored.

The water insertion into the S$_2$ open cubane has gained substantial support most recently~\cite{ibrahim2020untangling, hussein2021structural, allgower2022molecular, okamoto2021proton}. In our simulations, this is subject to the path-2 free of a transition to S$_2$-b closed cubane isomer. The path-2 starts with the proton removal from W1 under the hole polarization field, resulting in the Mn spin configuration change to $\uparrow\uparrow\downarrow\uparrow$ in the simulation for both simple model and the realistic hole at Ty161. Subsequently, W3 is inserted as a substrate water between Mn$_1$ and O$_5$. During the S$_2\rightarrow$S$_3$, it is noteworthy to show that the amino acid Glu189 moves away from Ca and Mn$_1$-Mn$_4$ distance elongates, as also reported in Ref.~\cite{ibrahim2020untangling} with the observation around 50$\sim$150$\mu$s. These structural changes are closely related to the event of the W3 insertion. With some path dependence of the insertion,  the W3 can be inserted in two forms: either as a spontaneous dissociation of OH$^-$ and H$^+$ or as the H$_2$O, as shown on the branch (I) and (II) of path-2 in Fig.~\ref{fig5} where the intermediate states are denoted by S$_2$-a-w3 on path-2 (I) and S$_2$-a-w3$^\prime$ on path-2 (II),  respectively. Along  the  path-2 (I) from  the diagram (C), one of  protons  from W3, denoted as Ha or Hb, transfers through W2 to W1. With the Ha transfer by option, the spin configuration of Mn changes to $\uparrow\downarrow\uparrow\downarrow$, corresponding to the path from  diagram (D) to (E) of path-2 (I) in Fig.~\ref{fig5}. In the case of the Hb transfer, the Mn spin configuration changes  through $\uparrow\downarrow\downarrow\uparrow$ to $\uparrow\uparrow\downarrow\uparrow$, which is the path from  the diagram (G) and (H) of path-2 (I). In the subsequent simulations, as the moving electron approaches the hole, the Mn spin configuration goes through $\uparrow\downarrow\downarrow\uparrow$, $\uparrow\downarrow\uparrow\uparrow$  and $\uparrow\downarrow\uparrow\downarrow$ to $\uparrow\uparrow\downarrow\uparrow$ in the path to diagram (F) and undergoes $\uparrow\downarrow\downarrow\uparrow$   to $\uparrow\uparrow\downarrow\uparrow$  in the path to diagram (I). Here, the $\uparrow\uparrow\downarrow\uparrow$ of S$_3$-a$^{\prime\prime}$ in the diagram (I) is  a Mn spin configuration  confirmed by the $^{55}$Mn hyperfine coupling constants in the S$_3$ state~\cite{krewald2015metal, cox2014electronic}. Besides this path with the Hb removal, in our simulation another isomer is predicted with the same spin configuration induced by the Ha removal.  Along the path-2 (II) with the intact H$_2$O insertion at the state S$_2$-a-w3$^\prime$, the proton Ha then  transfers to W1. At this step, the constraint calculations are performed to obtain the Mn spin configuration of the low-lying state amongst all possible intermediate states. It is the intermediate state S$_2$-a-w1$^{\prime\prime}$ with the  spin configuration $\uparrow\downarrow\downarrow\uparrow$ that  evolves to an energetically favorable state with the spin configuration $\uparrow\uparrow\uparrow\downarrow$, next to the subsequent electron transfer. Note that this Mn spin configuration of the state S$_3$-a in the diagram (L) is consistent with that  reported spin of S$_3$ in Refs~\cite{krewald2015metal,krewald2016spin}.

It is necessary to mention that the final-state geometric and electronic structure together with its spin configuration in path-2  is rather dependent on the intermediate state prior to the electron transfer. This dependence is largely ascribed to a rather flat potential energy surface of the intermediate states in a multi-step evolution to the  S$_3$ state.  In addition, the multiple spin flips between the low and high spin configurations in path-2 may imply that the spin exchange interaction plays an active part in the present functional and thus a comprehensive consideration of the   exchange interaction seems to be necessary in more accurate functionals. Though recent experiments have made remarkable achievements in discriminating the main evolution stages~\cite{ibrahim2020untangling, hussein2021structural, allgower2022molecular, okamoto2021proton}, our predictions that reveal the extensive details and information of the dynamic evolution, especially those belonging to the intermediate states, remain to be verified  by experiments in the future.

In this work, we are mainly dedicated to the specification and characterization of the polarization effect on the OEC evolution with a medium-size model that does not include the Tyr161 in the density functional simulations. Some dynamic aspect associated with the Tyr161 structure may thus be missing. For instance, the switch effect of the hydrogen bond connecting the O atom of Tyr161, which is predicted in the Tyr161 oxidation~\cite{nakamura2020pivotal}, is absent in the simulations.  While we performed the simulations of the polarization effect by putting an equivalent positive point charge at Tyr161, the subsequent impact of tyr161 oxidation is actually included phenomenologically on the water orientation and the OEC system. Yet, expanding our model would be necessary to include more dynamic content in future.

\section{Summary}

We have investigated the static properties and the dynamic evolution of the oxygen evolving complex from the state S$_0$ to S$_3$ driven by  the polarization field of the photoproduced hole using the density functional method. The simulation with appropriate modelling of the polarization field shows that the Mn spin flips occur with relevant structural changes in key Mn-O distances  which can serve as the precursor of losing electron from relevant Mn atoms. Constrained by the geometric and electronic structure of various states, the emphasis was put on the dynamic evolution traced by the spin marker that points to the correct spin configuration of the next state. In the processes of S$_0$$\rightarrow$S$_2$, the corresponding spin flips, following the electron and proton transfers, are displayed dynamically under the evolving polarization field. For the multi-step evolution of  S$_2$$\rightarrow$S$_3$ with water insertion, we reconstructed the evolution pathways, consistent with the main intermediate stages captured by the observations. In particular, a spontaneous  water dissociation between Mn$_4$ and O$_5$ was observed with the new finding of the two final-state isomers with the same spin configuration.  Our study paves a way to uncover the animated evolutions of the Kok's cycle with spin flips and lays a solid ground for revealing the mechanism of dioxygen generation by reestablishing the elusive S$_4$ state guided by the spin flips.

\section*{Acknowledgement}
We thank  Profs. Xi-Jun Qiu,  Han-Sen Gu and Dr. Shi-Jun Yuan for some early instructive discussions. This work was supported in part by the National Natural Science Foundation of China under grants No. 11775049 and No. 12375112 and the China Postdoctoral Science Foundation under grant No. 2021M690627. The Big Data Computing Center of Southeast University is acknowledged for providing the facility support on the partial numerical calculations of this work.

\bibliography{ref}

\end{document}